\documentclass[runningheads]{llncs}

\usepackage{tikz}
\usetikzlibrary{arrows,positioning,shapes.geometric}
\usepackage{verbatim}
\usepackage{pgfplots}
\usepackage{colortbl}
\usepackage{amssymb}
\usepackage{multirow}
\usepackage{mathtools}
\usepackage{amsmath}
\usepackage{comment}
\usepackage{enumitem}
\usepackage{textcomp}
\usepackage{subcaption}
\usepackage{perpage} 
\MakePerPage{footnote} 

\begin{document}
\title{Improving SIEM for Critical SCADA Water Infrastructures Using Machine Learning}
\titlerunning{Improving SIEM for Critical SCADA Water Infrastructures}
%
%
\author{Hanan Hindy\inst{1} \and 
David Brosset\inst{2} \and 
Ethan Bayne\inst{1}  \and \\
Amar Seeam\inst{3}  \and
Xavier Bellekens \inst{1}}
\authorrunning{H. Hindy et al.}
%
\institute{Division of Cyber Security, Abertay University, Dundee, Scotland, UK 
\email{\{1704847,e.bayne,x.bellekens\}@abertay.ac.uk}\\
\and Naval Academy Research Institute, France \\
\email{david.brosset@ecole-navale.fr} \and
Department of Computer Science, Middlesex University, Mauritius \\
\email{a.seeam@mdx.ac.mu}}
\maketitle              
\begin{abstract}
Network Control Systems (NAC) have been used in many industrial processes. 
They aim to reduce the human factor burden and efficiently handle the complex process and communication of those systems.
Supervisory control and data acquisition  (SCADA) systems are used in industrial, infrastructure and facility processes (e.g. manufacturing, fabrication, oil and water pipelines, building ventilation, etc.)
Like other Internet of Things (IoT) implementations, SCADA systems are vulnerable to cyber-attacks, therefore, a robust anomaly detection is a major requirement. 
However, having an accurate anomaly detection system is not an easy task, due to the difficulty to differentiate between cyber-attacks and system internal failures (e.g. hardware failures). 
In this paper, we present a model that detects anomaly events in a water system controlled by SCADA. 
Six Machine Learning techniques have been used in building and evaluating the model. 
The model classifies different anomaly events including hardware failures (e.g. sensor failures), sabotage and cyber-attacks (e.g. DoS and Spoofing).
Unlike other detection systems, our proposed work focuses on notifying the operator when an  anomaly occurs with a probability of the event occurring. This additional information helps in accelerating the mitigation process.
The model is trained and tested using a real-world dataset.

\end{abstract}

\newcolumntype{C}[1]{>{\centering\arraybackslash}m{#1}}
\newcolumntype{L}[1]{>{\begin{math}}c<{\end{math}}}
\setlength{\belowcaptionskip}{-5pt}

\begin{figure}[t]
	\centering
    \includegraphics[width=0.85\textwidth]{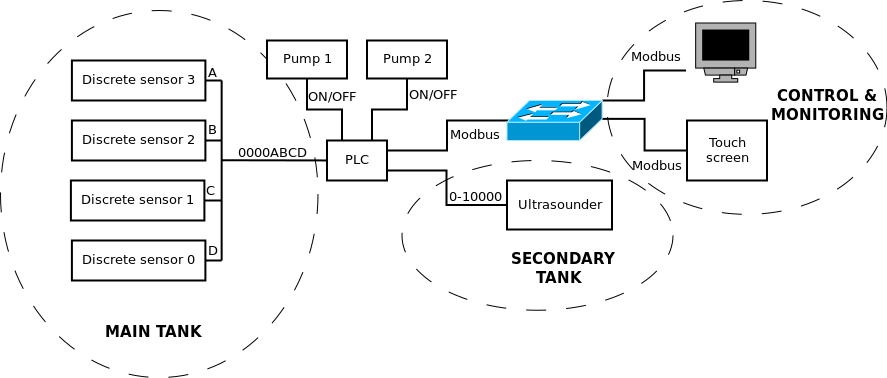}
    \caption{Network Architecture}
    \label{fig:networkArchitecture}
\end{figure}

\section{Introduction}
An increasing number of networked control systems are being deployed for the monitoring, control and response of physical infrastructure. Deprecated infrastructure is being replaced by SCADA systems that are compatible with IoT devices, enabling remote monitoring and control of previously isolated systems. Through this upgrade, industries are able to achieve higher reliability and flexibility of deployed systems. However, by enabling external internet access, industries are introducing the increased risk of cyber-attacks to potentially critical infrastructure~\cite{ten2010cybersecurity}.

Vulnerabilities in critical infrastructure protection (CIP) infrastructure is due to five factors. The first is the \textit{lack of open protocols}. Current programmable logic controllers (PLCs) are often controlled through proprietary network protocols that have not been scrutinized by security professionals and act as a black box, hence increasing the attack surface~\cite{calderon2018integration}. The second is due to the \textit{lack of segmentation} in network infrastructure~\cite{jiang2017research}. Current PLCs are connected to control networks for the data to be analysed in real time by higher management. Whilst this may increase profit margins and decrease the reaction time to market changes, the lack of segmentation allows for transversal network attacks. The third one is due to the \textit{number of off-the-shelf equipment} being set up in the network~\cite{bujari2018standards}. By including off-the-shelf hardware, the attack surface increases, and introduce other potentially weak links (i.e. Entry points). The fourth one is due to the \textit{lack of training received} by the operators and the inability of the operator to distinguish between incidents and cyber-attacks~\cite{bellekens2015cyber,ahmed2016scada}. Finally, the fifth one is due to the exponential growth of organized cyber-crime and nation state-sponsored cyber-warfare to destabilize countries~\cite{jensen2002computer,tan2018cyber}. 

Critical infrastructure network Security is intrinsically different from computer networks as the interaction between the nodes is done in real time at a physical level. Numerous efforts have been made to apply current cyber-security solutions to CIP networks, however, the solutions proposed are often not suited for the infrastructure and do not consider the underlying risks posed by compromised sensor data~\cite{gupta2017editorial}. As a results, despite the advances made in the design and implementation of cyber-security solutions, there is little research being made with the goal of improving SIEM and increase the resilience of CIP against cyber-incidents. In recent years, the number of attacks against critical infrastructure has significantly increased. The Stuxnet malware is an example of a high-profile SCADA attack, which was first discovered in an Iranian power plant in 2010. Stuxnet was used to infect and reprogramme the PLCs while hiding the changes made using a custom-made PLC rootkit \cite{5772960}. Another example is the Maroochy attack against a sewage system, causing 800.000L of sewer waste to be released in waterways and parks. The Maroochy attack was perpetrated by an insider, taking advantage of insecure communication between the pumping station and the central SCADA control system \cite{doi:10.1177/0096340213501372}. The Ukrainian power grid attack is another example of a large SCADA attack, where 225,000 people were affected by the loss of power. The attacker managed to switch off 30 substations for over 6 hours. In total, over 74 MWh of electricity was not supplied to the energy grid, representing a total of 0.015\% of the daily electricity consumption of the country. The attack was later attributed to "Sandworm", a Russian group of hackers known for advanced persistent threats \cite{case2016analysis}. These attacks demonstrate that when attacking a CIP, attackers often study the system extensively to perpetrate the attack as stealthily as possible, tailoring their strategies towards the particular system.

In this manuscript, we aim at improving Security Information and Event Management (SIEM) for critical infrastructure using machine learning to identify patterns in the data reported by PLCs in a water system controlled by SCADA. The contributions of this paper are as follows:
\begin{itemize}
\item We provide a new SIEM methodology that leverage sensor data and the event-driven nature of cyber-physical systems. 
\item We categorize the attacks in 14 categories, based on the data emerging from the dataset. Out prototype features 3 experiments using machine learning algorithms for the detection of the said anomalies.
\item We conduct a thorough evaluation of SIEM performances through a operational scenarios in the third experiment. The two best predictions are proposed to the operator allowing him to alleviate cyber-attacks when an event is detected. The anomaly detection provides an accuracy of 95.64\% with an 85\% confidence. 
\end{itemize}

The remainder of this paper is organized as follows; Section~\ref{sec:experimentalsetup} discusses equipment used to generate the dataset. Section~\ref{sec:dataset} provides an overview of the dataset, the scenarios and the features, while Section~\ref{sec:results} presents the results obtained by applying machine learning technique to differentiate between different scenarios. Section~\ref{sec:discussionandlimitations} discusses the results and limitations discovered. Section~\ref{sec:relatedwork} presents related work in this field. The paper ends by concluding our findings in Section~\ref{sec:conclusion}.
\section{Experimental Set-up}
\label{sec:experimentalsetup}
In this section we describe the architecture of SCADA controlled critical infrastructure (CI) used to gather the dataset described in Section~\ref{sec:dataset}. Current CI systems are interconnected and are increasingly prone to faulty operations and cyber-attacks. Risks present are caused by the fragmentation of technology incorporated into the CI, vulnerabilities within hardware and software components and through physical tampering of equipment by malicious actors. The design of the CI presented in this research represents a real world, real-time system that is capable of working under normal and abnormal conditions. Additionally, the presented system can be exposed to the aforementioned vulnerabilities. 

Figure~\ref{fig:networkArchitecture} provides a high level overview of the network of the system. The system is composed of two tanks--- the main tank and a secondary tank. Each tank can contain either fuel or water and can be set to two distinct modes--- acting either as a distributor or as storage. The main and secondary tank has a capacity of 9 and 7 litres respectively. The main tank is composed of four sensors connected to a PLC. The sensors are then connected to both Pump1 and Pump2. Both pumps control the flow of water between the main tank and the secondary tank. The main tank utilizes the four sensors to measure the level of liquid, whilst the secondary tank monitors fluid volume with an ultrasound sensor. Data gathered from the sensors are transferred to the control and monitoring network using the Modbus protocol.

\subsection{Modbus Protocol}
The Modbus protocol operates at the application layer of the open systems interconnection (OSI) model. It enables communication between interconnected network nodes based on a request and reply methodology. The protocol requires little processing overhead, which makes it a sensible candidate for communications between PLCs, Sensors, or Remote Terminal Units (RTUs). 

The Modbus protocol works independently of protocols implemented in other layers of the OSI model, hence, it can be used both on routable network or used for serial communications~\cite{huitsing2008attack}.

The data being stored by the Modbus protocol in slave devices can be categorized in four different ways as listed below:
\begin{itemize}
\item \textbf{Discrete Input} Read-Only Access, it provides Physical I/O
\item \textbf{Input Register} Read-Only Access, it provides Physical I/O
\item \textbf{Holding Register} Read-Only Access, it provides Read-Write Data 
\item \textbf{Coil} Read-Write Access, it provides application data
\end{itemize}

Each table can contain up to 9,999 values, however, some devices can allow up to 65,536 addresses across all tables. Each vendor has its own specification of the Modbus data tables, and the set-up often requires the operator to read the vendor-specific documentation. 

\begin{figure}[t]
	\centering
    \includegraphics[width=0.95\textwidth]{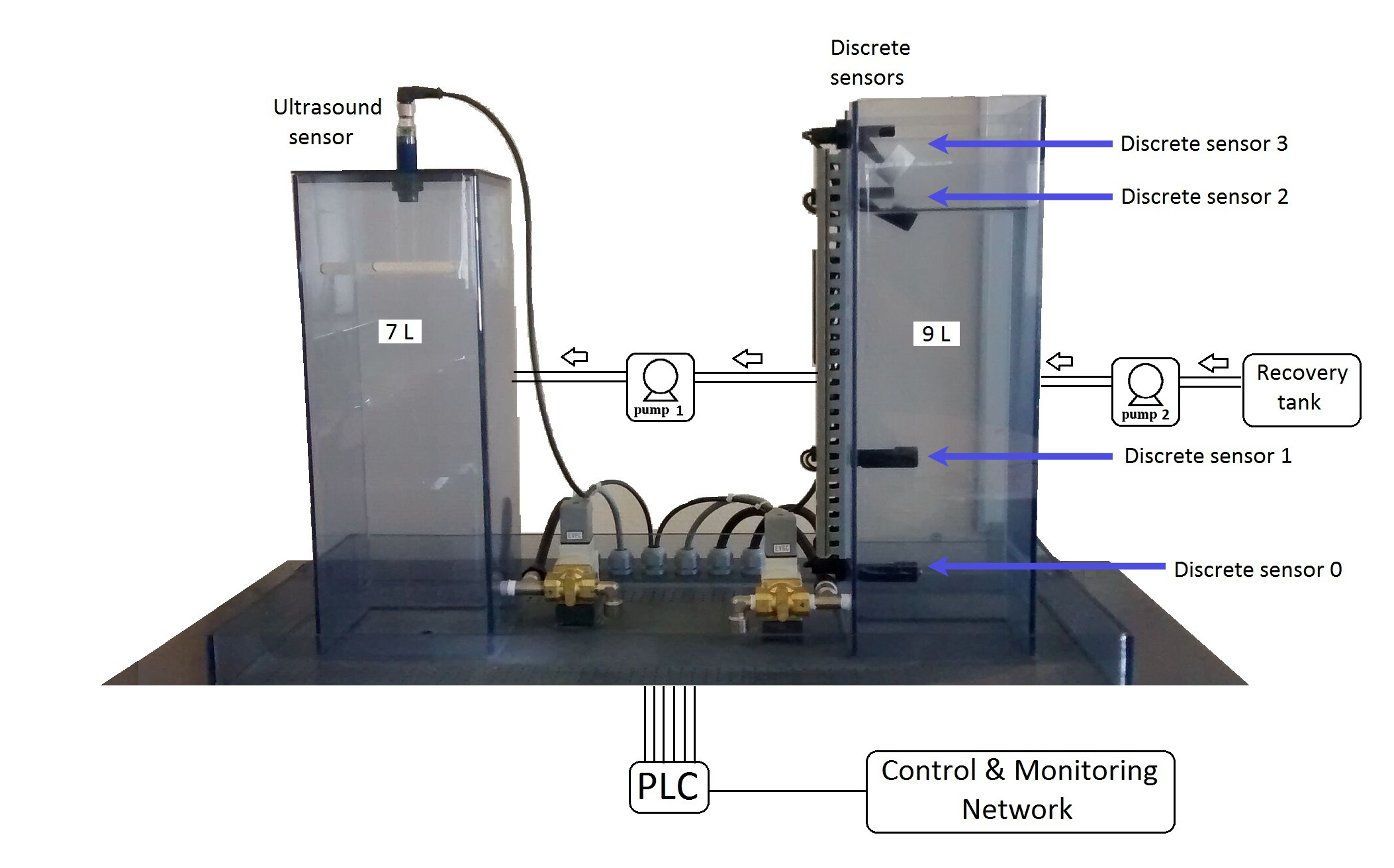}
    \caption{System Architecture}
    \label{fig:systemarchitecture}
\end{figure}

\subsection{Sensors and PLC}
As shown in Figure~\ref{fig:networkArchitecture} the system is composed of a single PLC. The PLC is a compact Base Twido PLC, with 40 discrete I/O, 24 discrete inputs, and 14 relays. The system is also composed of a Modicon M238 logic controller. The system can be sent instructions through the touch screen or via a remote system connected to the network. 

The system is also composed of five sensors as shown in Figure~\ref{fig:networkArchitecture} . Four discrete level sensors are located in the main tank. The first sensor (S0) indicates a low level in the tank (1.25L), the second sensor (S1) indicates a measure of less than 3.35L, the third sensor (S2), indicates a level of 8L while the last sensor (S3) indicates a full tank measure (9L).

The secondary tank is fitted with a single ultrasound sensor. The ultrasound sensor is a Schneider-Electric cylindrical M18 ultrasonic sensor. The sensor is fitted at the top of the tank and is used to measure the distance of the liquid surface to the top of the tank. The sensor can also be used to detect the presence or absence of liquid in the tank. 

\subsection{Operation Mode}
The system is controlled automatically through the PLC to avoid the spillage of liquid stored in tanks. Figure~\ref{fig:systemarchitecture} illustrates the physical system. For the purpose of this research, the primary tank is filled from a recovery tank, however, the recovery tank in our system can be substituted by other liquid source, such as a river, a fuel line, etc.

Once the primary tank is filled from the recovery tank using pump2, the PLC activates Pump1 to transfer liquid to the secondary tank to avoid spillage. 

To simulate a constant liquid consumption from the tanks, valves at the bottom of both the primary and the secondary tanks can be opened. The PLC monitors tank volumes by reading all sensor data registers at 0.1 second intervals. The PLC will automatically refuel the primary tank using Pump 2 when the volume of liquid goes below 1.25L. Similarly, Pump2 will be instructed to stop when a total volume of 9L is reached. The secondary tank will be refuelled when ultrasound sensors detect the liquid level to be below 2.1L. Furthermore, the PLC will only deactivate pump 1 once the ultrasound sensor detects that the secondary tank has reached a level of 6.3L.

\section{Dataset}
\label{sec:dataset}
This section describes the dataset gathered by the IC and the scenarios recorded. The dataset used in this manuscript has been published separately in a dataset publication~\cite{RefWorks:doc:5ad0a812e4b0e18303dc2c3e}.

The CI presented in section~\ref{sec:experimentalsetup} was used to gather data for the dataset and outputted into CSV format. The dataset comprises actuator and sensor readings that the PLC recorded periodically at 0.1 second intervals. Within the data collected, PLC registers 2 through 4 provided output data describing the state of the system that was used for analysis.

\begin{table}[tb]
\caption{Registers Extracted Bits Representation}
    \begin{minipage}{.5\textwidth}
      \centering
        \begin{tabular}{| C{0.2\textwidth} | C{0.2\textwidth} | C{0.5\textwidth}|} 
        \hline
        \textbf{Reg. No.} &\textbf{Bit No.} & \textbf{Value}  \\
        \hline 
        \multirow{4}{*}{2} & 4 & Discrete Sensor 3 \\
        \cline{2-3}
        & 5 & Discrete Sensor 2 \\
        \cline{2-3}
        & 6 & Discrete Sensor 1 \\
        \cline{2-3}
        & 7 & Discrete Sensor 0 \\
        \hline
        4 & 16-bits & Depth Sensor \\
        \hline
        \end{tabular}
    \end{minipage}%
    \begin{minipage}{.5\textwidth}
      \centering
       \begin{tabular}{| C{0.2\textwidth} | C{0.2\textwidth} | C{0.5\textwidth}|}
        \hline
        \textbf{Reg. No.} &\textbf{Bit No.} & \textbf{Value}  \\
        \hline 
        \multirow{4}{*}{3} & 0 & Pump 2 \\
        \cline{2-3}
        & 1 & Pump 1 \\
        \cline{2-3}
        & 5 & Pump 2 Valve \\
        \cline{2-3}
        & 4 & Pump 1 Valve\\
        \hline
        \end{tabular}
    \end{minipage} 
    \label{tab:bits}
\end{table}

Table~\ref{tab:bits} provides an overview of the different registers used by the PLC. As shown, Register 2, provides the bits indicating the binary state of the discrete sensors. A population count can be done on the register to retrieve the state of each sensor separately. Register 3 contains the state of the pump, either as active or inactive, while Register 4 contains the step value from 0 to 10,000 of the ultrasound sensors (e.g. Step 3,000 represents 2.1L of liquid in the tank). 

\subsection{Scenarios}
The dataset consists of 14 different scenarios as shown in Table~\ref{tab:dataset-scenarios}. Each scenario covers one of 5 operational scenarios representing the potential threat (i.e. sabotage, accident, breakdown, or cyber-attack) as well as 6 affected components. The affected components are system components that are directly affected by the anomaly. 

The recorded data is organized in 15 different CSV files of variable duration based on the type of incident recorded (i.e. a sabotage incident may take less time than a distributed denial of service). 

\begin{table}[bt]
 \caption{Dataset Scenarios, Operational Scenarios and Affected Components}
\begin{tabular}{| C{0.05\textwidth} | C{0.25\textwidth} | C{0.25\textwidth} | C{0.25\textwidth} | C{0.15\textwidth} |} 
\hline
 & \textbf{Scenario} &\textbf{ Affected Component} & \textbf{Operational Scenario} & \textbf{No. of instances}  \\
\hline 
1 & Normal & None & Normal & 5519 \\
\hline
2 & Plastic bag & \multirow{6}{*}{Ultrasound Sensor} & Accident/Sabotage & 10549 \\
\cline{1-2} \cline{4-5}
3 & Blocked measure 1 & &  \multirow{2}{*}{Breakdown/Sabotage} & 226 \\
\cline{1-2} \cline{5-5}
4 & Blocked measure 2 & & & 144  \\
\cline{1-2} \cline{4-5}
5 & Floating objects in main tank (2 objects) & & \multirow{2}{*}{Accident/Sabotage} & 854  \\
\cline{1-2} \cline{5-5}
6 & Floating objects in main tank (7 objects) & & & 733  \\
\cline{1-2} \cline{4-5}
7 & Humidity & & \multirow{3}{*}{Breakdown} & 157 \\
\cline{1-3}\cline{5-5}
8 & Discrete sensor failure & Discrete sensor 1 & & 1920  \\
\cline{1-3}\cline{5-5}
9 & Discrete sensor failure & Discrete sensor 2 &  & 5701 \\
\cline{1-5}
10 & Denial of service attack & \multirow{3}{*}{Network} & \multirow{2}{*}{Cyber-attack} & 307 \\
\cline{1-2} \cline{5-5}
11 & Spoofing & &  & 10130 \\
\cline{1-2} \cline{4-5}
12 & Wrong connection & & Breakdown/Sabotage & 6228 \\
\cline{1-5}
13 & Person hitting the tanks (low intensity) & \multirow{3}{*}{Whole subsystem	}  & \multirow{3}{*}{Sabotage}  & 347 \\
\cline{1-2}\cline{5-5}
14 & Person hitting the tanks (medium intensity) & &   & 281\\
\cline{1-2}\cline{5-5}
15 & Person hitting the tanks (high intensity) & & & 292  \\
\hline
\end{tabular}
 \label{tab:dataset-scenarios}
\end{table}

\subsection{Pre-processing}
This subsection highlights the pre-processing of the data obtained. The pre-processing is composed of six steps. 
\begin{enumerate}
\item Extract Instances
\end{enumerate}
Firstly, each scenario instances are extracted from the log file. An instance is represented by the recording of the register values at a specific time. 
Each log file has 10 rows per instance. Each row contains The Date, Time, the Register Number, and the Register Value of the PLC.

\begin{figure}[t]
    \centering
    \includegraphics[width=\textwidth, height=0.45\textheight]{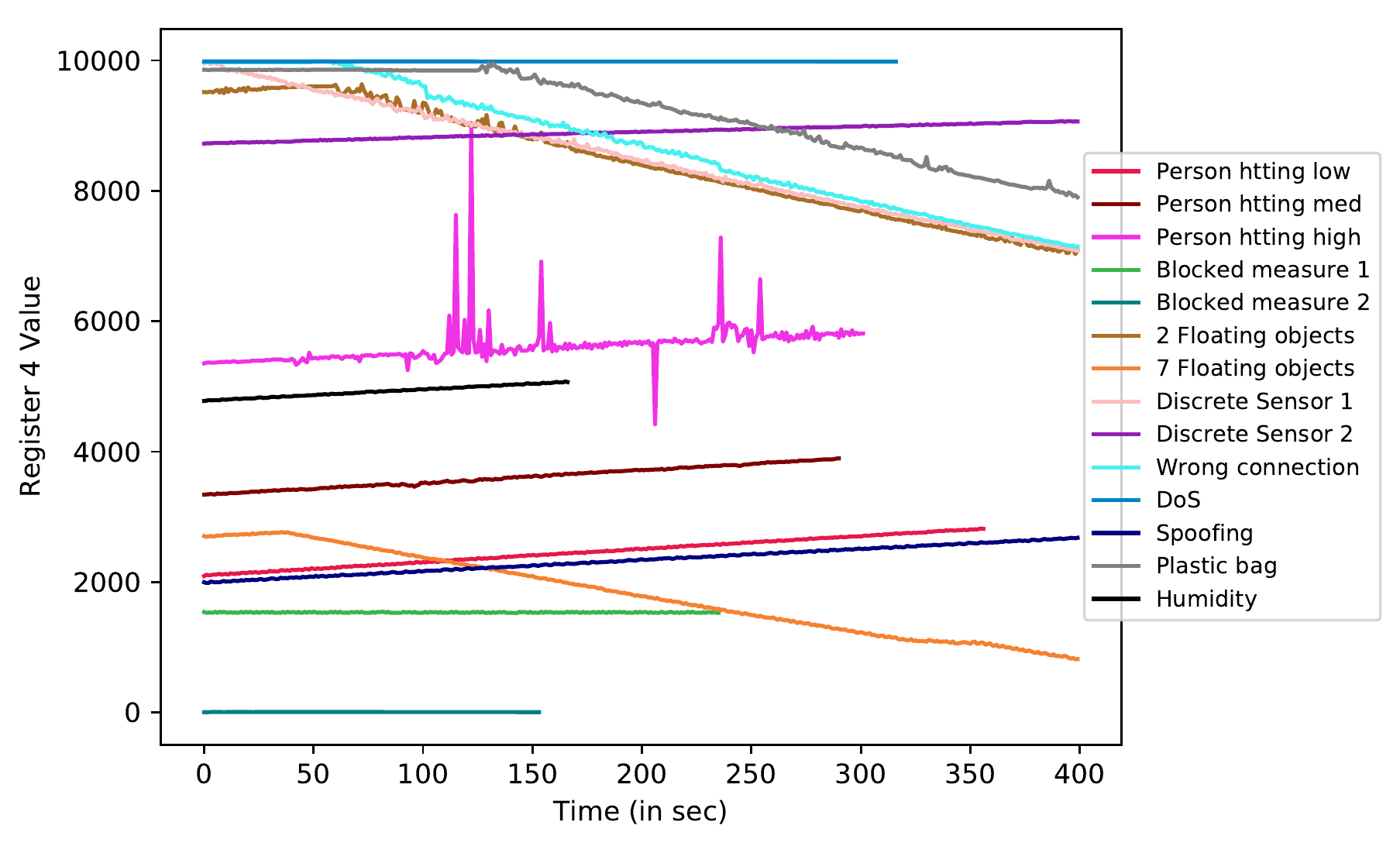}
    \caption{Rate of change of register 4 value over different scenarios}
    \label{fig:rate-of-change}
\end{figure}

\begin{enumerate}[resume]
\item Calculate rate of change of Register 4
\end{enumerate}
The value of Register 4 is the most significant. Its significance can be seen in the rate of change over time. Figure~\ref{fig:rate-of-change} shows the rate change in Register 4 value over 15 different scenarios.

For each instances, the rate of change is calculated over 10 time frames as expressed in Equation~\ref{rate-of-change}. 

\begin{equation}
\label{rate-of-change}
Rate\ of\ change_i = \dfrac{reg4_{i} - reg4_{i - 10} }{time_i - time_{i-10}}
\end{equation}

\begin{enumerate}[resume]
\item Apply Threshold
\end{enumerate}
Table~\ref{tab:dataset-scenarios} shows the number of recorded instances per scenario. As shown in the table, the events are not statistically distributed over the scenarios. Therefore, the most instances scenario will bias the classification model output. A threshold is applied to take only the first $N$ instances of each file.

\begin{enumerate}[resume]
\item Serialization \\ A single file is needed for the training process. All Instances are, therefore, serialized. 
\item Normalization 
\item Split 
\end{enumerate}

Finally, the data is normalized and split into 80\% training and 20\% testing sets. 

\section{Experiments and Results}
\label{sec:results}
This section outlines three experiments conducted and how accurately threats could be identified. 

The aim of the experiments is to alert, and provide the operator with the most likely affected components, in order to decrease the time to apply the corrective action. 

Six machine learning techniques are used for the classification models, namely, Logistic Regression~(LR) \, Gaussian Na{\"i}ve Bayes~(NB) , k-Nearest Neighbours~(k-NN) , Support Vector Machine~(SVM) , Decision Trees~(DT)  and Random Forests~(RF) \cite{RefWorks:doc:5af4661ee4b0dfb887b46a75}~\cite{RefWorks:doc:5af46286e4b02abf496e090c}~\cite{RefWorks:doc:5af4651ee4b02dfcb38d5a2b} \cite{RefWorks:doc:5af4638ae4b0f7bd1fabb685}~\cite{RefWorks:doc:5af464f7e4b0cfc1f4ac242d}~\cite{RefWorks:doc:5af4642de4b04303c30e488f}.

\subsection{Parameters}
All experiments are conducted using the following computation parameters:
\begin{itemize}
\item Training Set : Testing Set = 80\% : 20\%
\item Evaluation: The overall accuracy is calculated as follows:
    	\begin{align*}
			Overall Accuracy = \dfrac{TP + TN}{TP + TN + FP + FN} 
        \end{align*}
        Where: 
        \begin{itemize}
          \item True Positive (TP): Number of anomaly instances which scenario is correctly detected
          \item True Negative (TN): Number of normal instances which are correctly detected 
          \item False Positive (FP): Number of normal instances which are detected as one of the anomaly scenarios 
          \item False Negative (FN): Number of anomaly instances which are detected as normal scenarios 
	\end{itemize}
\end{itemize}

\subsection{First Experiment}
\begin{figure}[tb]
  \centering
  \begin{minipage}[b]{0.48\textwidth}
    \includegraphics[width=\textwidth]{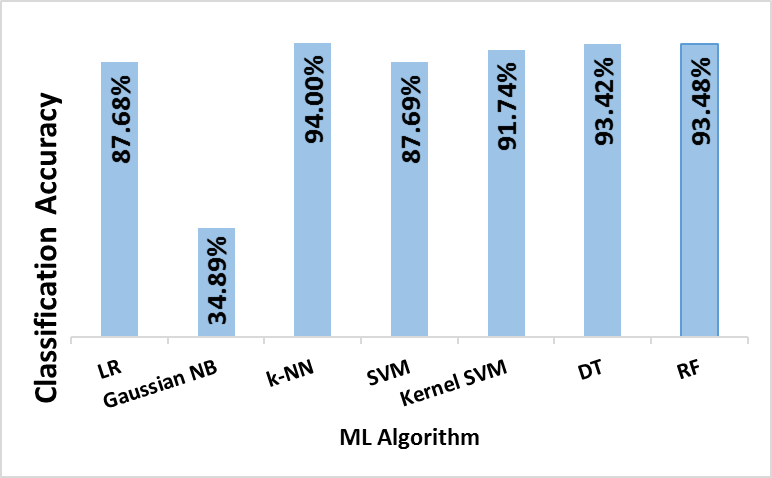}
    \caption{Anomaly Classification Accuracy}
    \label{fig:res-anomaly}
  \end{minipage}
  \hfill
  \begin{minipage}[b]{0.48\textwidth}
   \includegraphics[width=\textwidth]{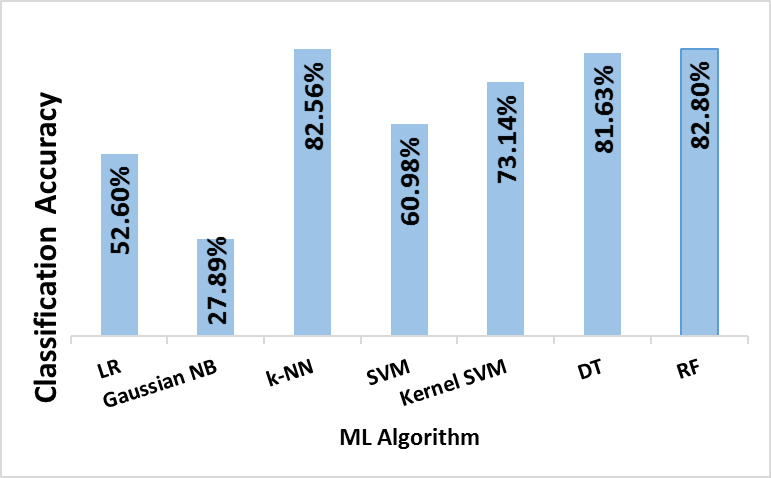}
    \caption{Affected Component\\Classification Accuracy}
    \label{fig:res-affected-component}
  \end{minipage}
\end{figure}

The aim of the first experiment is to alert the user when an anomaly occurs without specifying the associated scenario. The aim is to provide the operator of the CI with a binary output.

Figure~\ref{fig:res-anomaly} shows the classification results of the six machine learning algorithms applied to data. The algorithms provide a binary output. As shown, the highest accuracies reached are 93.42\%, 93.48\% and 94\% using DT, RF and k-NN respectively.

Providing an alert that an anomaly exists in the data recorded by the sensors is considered important, however, as the alert is provided in a binary fashion, the operator is -- in this case -- unable to identify the anomaly at first sight and take corrective action. To this end, a second experiment was established with the aim of proving more information to the operator. 

\subsection{Second Experiment }
The second experiment aims at giving the operator capital information about the anomaly. The alert in this case includes the affected component. The model classifies what components are affected by the five components present in the CI (i.e. Network, Discrete Sensor 1, Discrete Sensor 2, Ultrasound Sensor and Whole System) as well as the 'None' affected case. 

Figure~\ref{fig:res-affected-component} shows the classification results of the different machine learning algorithms. The highest accuracies reached are 82.56\% and 82.8\% using k-NN and RF respectively. The result shows a trade-off between the binary classification offered in the first experiment and by providing the operator with more detailed information. As shown trade-off limits, the accuracy of the system accounts for a large number of false positive, potentially misleading the operator during normal operations. 

\subsection{Third Experiment}
\begin{figure}[tb]
  \centering
  \begin{minipage}[b]{0.45\textwidth}
    \includegraphics[width=\textwidth]{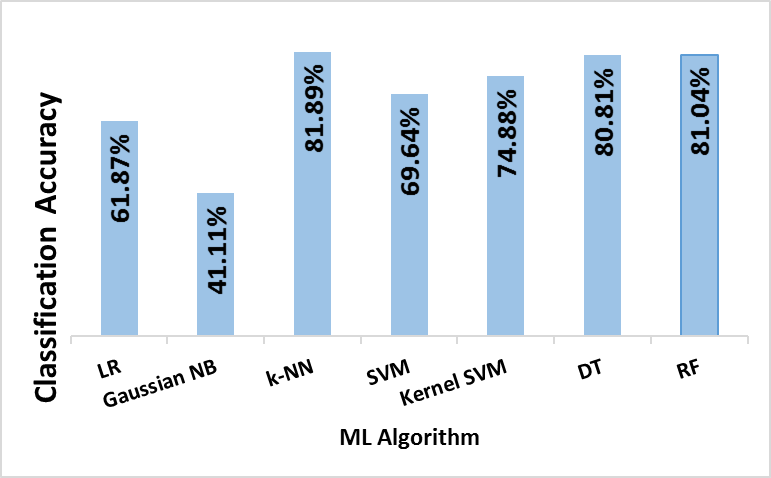}
    \caption{Scenarios Classification Accuracy. Trial One (Single scenario reported)}
    \label{fig:res-scenario-no-prob}
  \end{minipage}
  \hfill
  \begin{minipage}[b]{0.45\textwidth}
   \includegraphics[width=\textwidth]{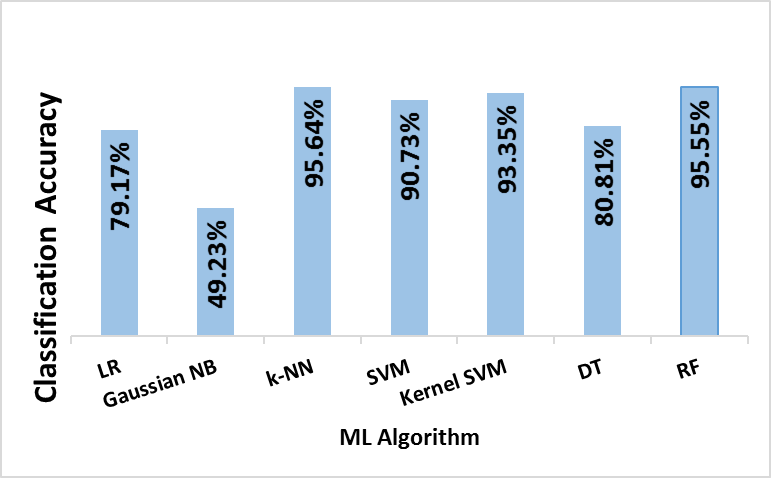}
    \caption{Scenarios Classification Accuracy (Alerting with two probable scenarios)}
    \label{fig:res-two-scenarios}
  \end{minipage}
\end{figure}

The third experiment aims at notifying the operator with an occurring scenario. To this end, the third experiment is divided in three operational trials.
\vspace{10mm} \\
\textbf{Trial One:}
\\
In the first trial, models are trained to classify the different scenarios. The operator is notified with the scenario occurring. Figure~\ref{fig:res-scenario-no-prob} shows the results of these trial. As shown, the accuracy is reduced and demonstrate a high false negative rate, with the highest accuracy only reaching 81.89\% 

To overcome this high false positive rate, the scenarios were reviewed and the following conclusions were made:
(a) The scenarios are co-related. 
(b) When models are trained to output the probability of an event  belonging to another of the 14 scenarios, a maximum of 4 scenarios have non-zero probabilities. 
For example, Table~\ref{tab:prob-classification-of-classes} shows the count of instances having 1, 2, 3 or 4 probable scenarios. 
In the second row of Table~\ref{tab:prob-classification-of-classes}, 61.64\% of the instances are classified to a single scenario, 31.51\% are classified to be one of 2 possible scenarios and only 6.85\% are classified to be one of 3 possible scenarios.

\begin{table} [tb]
\caption{Distribution of probabilistic classification of scenarios}
\begin{tabular}{| C{0.2\linewidth} | C{0.16\linewidth} | C{0.16\linewidth} |  C{0.16\linewidth} |  C{0.16\linewidth} | C{0.16\linewidth} |}
\hline

Algorithm & Maximum number of probable scenarios per instance & Number of instance with \textbf{1} probable scenario &   Number of instance with \textbf{2} probable scenarios & Number of instance with \textbf{3} probable scenarios & Number of instance with \textbf{4} probable scenarios \\
\hline
\multirow{2}{*}{DT} & \multirow{2}{*}{1} & 3053 & - & - & -  \\
\cline{3-6}
& & 100\% & - & - & - \\
\hline
\multirow{2}{*}{k-NN} & \multirow{2}{*}{3} & 1882 & 962 & 209 & - \\
\cline{3-6}
& & 61.64\% & 31.51\% & 6.85\% & - \\
\hline
\multirow{2}{*}{RF} & \multirow{2}{*}{4} & 1737 & 1004 & 297 & 15  \\
\cline{3-6}
& & 56.9\% & 32.89\% & 9.71\% & 0.5\% \\
\hline
\end{tabular}
\label{tab:prob-classification-of-classes}
\end{table}

As a result of this analysis, trial two to four are based on reporting two probable scenarios to the operator, reducing the uncertainty of our approach.
\vspace{5mm} \\
\textbf{Trial Two:}
\\
In the second trial, the operator is notified with two potential attack scenarios instead of a single one. The two scenarios are the highest probabilities provided by the algorithms. Figure~\ref{fig:res-two-scenarios} demonstrates that this technique increased the accuracy to reach 95.55\% and 95.64\% using RF and k-NN receptively. By providing the operator with a probability in the attack scenario he is able to act accordingly and alleviate the attack, hence reducing the overall response time needed.

Table~\ref{tab:scenarios-corelation} shows an example of correctly classified instances when two probable scenarios are considered. The numbers are calculated in reference to the k-NN classifier. 
For example, in the first row, '2 Floating Objects' scenario misclassified instances are shown . 53 are misclassified as 'Plastic Bag' sabotage. However, 48 of them can be correctly classified by considering the second probable scenario.
In the same manner, in Table~\ref{tab:scenarios-corelation} row 4, 85 instances of the 'Plastic Bag' scenario are misclassified to be '2 Floating Objects'~(38 instances), 'Sensor Failure'~(24 instances), 'Spoofing'~(18 instances), 'Normal'~(3 instances) and 'Wrong Connection'(2 instances). 75 can be correctly classified with the second probable scenario in consideration. 

\begin{table}[tb]
 \caption{Co-relation of scenarios that are misclassified based on highest probable scenario and correct with the second probable one (Calculated based on k-NN experiment)}
\centering
\resizebox{\textwidth}{!}{
\begin{tabular}{| C{0.1\textwidth} | C{0.1\textwidth} | C{0.1\textwidth} | C{0.1\textwidth} | C{0.1\textwidth} | C{0.1\textwidth} | C{0.1\textwidth} | C{0.1\textwidth} | C{0.1\textwidth} | C{0.1\textwidth} | } 
\hline  
 & \multicolumn{2}{|c|}{Instances count} & \multicolumn{7}{|c|}{Scenario (Y)  The count of instances classified}\\
 & \multicolumn{2}{|c|}{where X} & \multicolumn{7}{|c|}{as Y while the correct is X}\\

\cline{2-10}
Scenario (X) & \textbf{Is Not} 1\textsuperscript{st} Scenario & \textbf{Is} 2\textsuperscript{nd} Scenario & 2 Floating Objects & 7 Floating Objects & Normal & Plastic Bag & Sensor Failure & Spoofing & Wrong Connection  \\
\hline
2 Floating Objects & 53	& 48 & - & - & - & 53 & - & -	& - \\  
\hline
7 Floating Objects & 5 & 5 & -	& -	& -	& -	& -	& -	& 5 \\  
\hline
Normal & 80 & 58 & - & - & - & 1 & 68 & - & 11\\  
\hline
Plastic Bag & 85 & 75 & 38 & - & 3 & - & 24 & 18 &  2\\  
\hline
Sensor Failure & 183 & 137 & - & - & 75 & 53 &  - & 5 & 50 \\  
\hline
Spoofing & 45 & 25 & - & - & - & 37 &  8 & - & - \\  
\hline
Wrong Connection & 102 & 72 & - & 5 & 23 & 5 & 69 & - & - \\  
\hline
 \end{tabular}
 }
 \label{tab:scenarios-corelation}
\end{table}

\begin{figure}[tb]
 \centering
  \includegraphics[width=\textwidth]{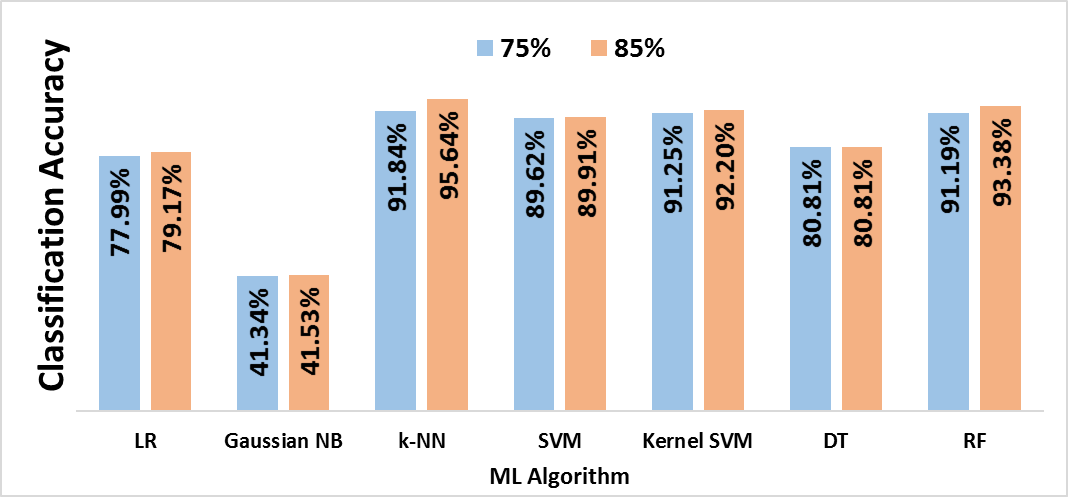}
    \caption{Scenarios Classification Accuracy (Alerting with one/two scenario(s) based on 75\% and 85\% confidence)}
    \label{fig:res-scenario-conf-75-85}
\end{figure}

While the operator is notified with two probable scenarios, it can also be misleading. For this reason our third attempt provides a confidence measure to the operator increasing his situational awareness.
\vspace{5mm} \\
\textbf{Trial Three and Four:}
\\
In the third and fourth trials, a single scenario is reported to the operator unless the probability is less than the confidence percentage.
Figure~\ref{fig:res-scenario-conf-75-85} shows the results for 75\% confidence. In the first case, a single scenario is reported if its probability is greater than or equal to 0.75, otherwise, if the threshold is reached a second scenario is reported as well. The accuracy provided is 91.84\%.
\\
This accuracy rises when using an 85\% confidence. Figure~\ref{fig:res-scenario-conf-75-85} shows a maximum accuracy of 95.64\% using k-NN. 

It can be seen from Figure~\ref{fig:res-two-scenarios} and~\ref{fig:res-scenario-conf-75-85} that DT accuracy is the same. This is due to the DT output, as a single scenario for each instance is available, therefore it is not possible to output two probable scenarios to the operator.
\\
Finally, the result shows that the fourth operational trial is the most convenient for the operator as a single scenario is provided, unless the threshold is reached. The accuracy provided by the fourth operational scenario also reaches 95.64\%, hence reducing the uncertainty of the attack currently at hand, and the ability of the operator to alleviate the attack.

\section{Discussion and Limitations}
\label{sec:discussionandlimitations}
In this section, we discuss the main takeaways based on the experiments evaluated in Section~\ref{sec:results}. Additionally, the limitations of our models are listed.
\\
\textbf{Key takeaways:}
\begin{itemize}
\item \textbf{Using high confidence:} Probabilities are considered to solve the trade-off between the co-related scenarios and reducing the amount of information provided to the operator (i.e. multiple scenarios). This implies reporting the probability of having a scenario occurring. For this piece of work, a confidence interval of 85\% is evidenced to provide the best results.

\item \textbf{Sensors location and architecture:} The position of the sensors and the architecture of the SCADA system affects the collected data and, consequently, affects the model detection accuracy. Moreover, the collection of data greatly affect the pre-processing stages, it is therefore important to gather data consistently. 

\item \textbf{Longer recording periods needed:} In order to increase the accuracy of the machine learning model, the operators should have the recordings of long periods of time for all sensor data. This helps have multiple instances over time and therefore, more instances for training. Moreover, it is recommended to have similar recording duration for the different scenarios. In the case of massively different recording lengths, files need to be serialised in order to avoid bias towards the scenario with the longest recording time. 

\item \textbf{Importance of scenarios:} The more recorded scenarios used for training, the more robust the detection model. With more scenarios, the model can learn variations in scenarios. Moreover, the operator will be provided with higher confidence of reported anomalies. 
\end{itemize}

\vspace{5mm}
\noindent
\textbf{Limitations:}
\begin{itemize}
\item{\textbf{Limited Number of Scenarios:}} The experiments were conducted using 15 log files which covered 14 scenarios. Any new scenario other than the ones covered in the log files will not likely be reported to the operator. 

\item{\textbf{Model Evaluation:}} Models are evaluated using six machine learning techniques. Hence, the results are limited to these algorithms.

\item{\textbf{Architecture:}} The data is limited to the scale of the CI used in this study. More complex architecture -- such as introducing more tanks and sensors -- may not yield the same anomaly detection accuracy.

\item{\textbf{Real-Time testing:}} The model is built and tested based on the dataset, however, testing the model real-time is essential to measure the performance.

\end{itemize}
\section{Related Work}
\label{sec:relatedwork}
The area of critical infrastructure protection and Security Information and Event Management has a rich history of books, deployments and lessons learned from both universities and organizations deploying tools and techniques to protect their environments. 

Robert Mitchell and Ing-Ray Chen~\cite{Mitchell:2014:SID:2597757.2542049} presented a survey on recent Cyber-Physical Systems (CPS) intrusion detection systems. They classified CPSs into two categories based on potential detection technique: (1) Knowledge-Based or Misuse Detection and (2) Behaviour-Based. While the survey provides a general overview of potential threats, it focuses essentially on the network traffic rather than on data provenance and data accuracy. This survey, however, provides the stepping stone of the field. In~\cite{amin2013cyber}~and~\cite{amin20132cyber} Amin~\textit{et al.} provide an overview a security threat assessment of a networked CI including different layers (i.e. supervisory and control networks). The manuscripts present a grey-box approach where the hacker possesses a certain level of knowledge of the system and is able to perpetrate a deception attack against the system in order to enable liquid spilling from the canal the IDS was tested on. Cheng~\textit{et al.}~\cite{cheng2017orpheus} provide a technique to alleviate control-oriented attacks, code-injection attacks or code-reuse attacks on embedded devices. The highlight the lack of existing mechanisms and present 'Orpheus', a program behaviour model, taking advantage of the event-driven nature of embedded devices controlling critical infrastructure. Mathur~\cite{mathur2018limits} discusses the limitations of the detecting a incidents in critical infrastructures by analysing processes. The manuscript refers to two methods, the first described is the CUSUM~\cite{cardenas2011attacks}. CUMSUM is a statistical method allowing to detect anomalies in time series, corresponding to a specified process. The technique requires two parameters to operate, the 'bias' and thethreshold'. The CUMSUM provides the cumulative sum of the deviation for the process measured. By plotting the predicted and observed state, the operator is able to identify the state of the facility and identify changes in behavior. The second methodology is based on State Entanglement (SE)~\cite{adepu2016distributed}. SE combines the states of multiple components of a system to construct a state space. The state space act as a blacklist, highlighting prohibited states during normal operations.  

\section{Conclusion and Future Work}
\label{sec:conclusion}

This work focused on building an anomaly detection and a SIEM tool for a SCADA water system. The model was evaluated using a real-world dataset covering 14 anomaly scenarios including normal system behavior. The presented scenarios covered a wide range of events, ranging from hardware failure to sabotage. Three experiments were conducted using 6 Machine Learning techniques.

The experiments varied based on the level of information reported to the operator. The First experiment allowed anomaly event to be reported to the operator as a binary output. While events where being detected as either anomalies or normal operation, the operator was unaware of the type of anomaly occurring. The second experiment reported the affected component, providing the operator with information relating to a single or multiple sensor data. Finally, the third experiment - which provided the best results - reported the anomaly with an accuracy level, helping the operator to take subsequent correcting steps. 

The overall evaluation showed that k-NN, Decision Tree and Random Forest outperformed 'Gaussian Na{\"i}ve Bayes, SVM and kernel SVM. Moreover, k-NN results demonstrated the highest accuracy amongst all algorithms in the three experiments. The experiments achieved the accuracy of \textbf{94\%} for the binary output and \textbf{95.64\%} for the scenarios classification.
As aforementioned, the scenarios are co-related, therefore, the operator - in the third experiment - is notified with the most probable scenarios/anomaly occurring. Moreover, a confidence level was used to provide the operator with the best information available.

To further enhance the detection accuracy, the following should be considered. Increasing the dataset instances to enhance the training process and building hybrid model to classify subgroups of events.

%
%
%
\bibliographystyle{splncs04}
\bibliography{bibliography}

\end{document}